\documentclass[authoryear,review,preprint,12pt]{elsarticle}
\usepackage[english]{babel}    %Le texte est en français
\usepackage[T1]{fontenc} %Pour les font postscript
\usepackage[cyr]{aeguill}%Police vectorielle, guillemets français
\usepackage{epsfig}%Gère l'importation de figures .jpeg
\usepackage{graphicx} %Gère l'importation de figure au format .eps de meilleure qualité
\usepackage{epstopdf}
\usepackage{lscape} %Pour utiliser le format paysage
\usepackage[utf8]{inputenc}
\usepackage{multirow}
\usepackage[breaklinks]{hyperref}
\usepackage[nottoc, notlof, notlot]{tocbibind} % Pour que la bibliographie apparaissent dans le sommaire
\usepackage{a4wide}
\usepackage{url}
\usepackage{color}
\usepackage{natbib}
\usepackage{lipsum}
\usepackage{multicol}
\usepackage{nopageno}
\usepackage{amsmath}
\usepackage{array}
\usepackage{amsthm}
\usepackage{numprint}
\usepackage{float}
\usepackage{amsfonts}
\newcommand{\argcosh}{\mathrm{argcosh}}

\newcommand{\subscript}[1]{\ensuremath{_{\textrm{#1}}}} 
\usepackage{numcompress}
\usepackage{fullpage}
\usepackage{empheq}
\usepackage{amssymb}
\usepackage{remreset}
\makeatletter \@removefromreset{footnote}{chapter} \makeatother
\usepackage {geometry}
\usepackage{subfigure}
\usepackage{listingsutf8}

\usepackage{wrapfig}
\begin{document}
\selectlanguage{english}

\begin{frontmatter}
\title{Theory for planetary exospheres: III. Radiation pressure effect on the Circular Restricted Three Body Problem and its implication on planetary atmospheres}
\author[icl,ups,irap]{A.~Beth\corref{cor1}}
\ead{arnaud.beth@gmail.com}
\author[ups,irap]{P.~Garnier\corref{cor2}}
\ead{pgarnier@irap.omp.eu}
\author[ups,irap]{D.~Toublanc}
\author[ups,irap]{I.~Dandouras}
\author[ups,irap]{C.~Mazelle}
\address[icl]{Department of Physics / SPAT, Imperial College London, United Kingdom}
\address[ups]{Université de Toulouse; UPS-OMP; IRAP; Toulouse, France}
\address[irap]{CNRS; IRAP; 9 Av. colonel Roche, BP 44346, F-31028 Toulouse cedex 4, France}

\cortext[cor1]{Principal corresponding author}
\cortext[cor2]{Corresponding author}
\begin{abstract}
The planetary exospheres are poorly known in their outer parts, since the neutral densities are low compared with the instruments detection capabilities. The exospheric models are thus often the main source of information at such high altitudes. We present a new way to take into account analytically the additional effect of the stellar radiation pressure on planetary exospheres. In a series of papers, we present with an Hamiltonian approach the effect of the radiation pressure on dynamical trajectories, density profiles and escaping thermal flux. Our work is a generalization of the study by \citet{Bishop1989}. In this third paper, we investigate the effect of the stellar radiation pressure on the Circular Restricted Three Body Problem (CR3BP), called also the photogravitational CR3BP, and its implication on the escape and the stability of planetary exospheres, especially for Hot Jupiters. In particular, we describe the transformation of the equipotentials and the location of the Lagrange points, and we provide a modified equation for the Hill sphere radius that includes the influence of the radiation pressure. Finally, an application to the hot Jupiter HD 209458b reveals the existence of a blow-off escape regime induced by the stellar radiation pressure. %The radiation pressure results from the resonant scattering of solar photons and induces a constant force, in first approximation, on the exosphere. In this paper, we present and give the complete exact solution for a species subject to planetary gravity and a constant radiation pressure with any initial conditions. Also, we introduce notations for the future work. %We apply the same approach as \citet{Bishop1989} and propose an extended version to the whole exosphere depending on two parameters: the distance from the planet and the Solar zenithal angle. With this model, we reproduce quite well the Earth observations of densities over 10 $R_{E}$ and highlight that the radiation pressure is responsible for exospheric asymmetries of $H$ density profiles in the higher exosphere.
\end{abstract}

\end{frontmatter}
\section{Introduction}
The exosphere is the upper layer of any planetary atmosphere: it is a quasi-collisionless medium where the particle trajectories are more dominated by gravity than by collisions. 
 Above the exobase, the lower limit of the exosphere, the Knudsen number {\it{Kn}} \citep{Ferziger1972} becomes large, collisions become scarce, the distribution function cannot be considered as maxwellian anymore and, gradually, the trajectories of particles are essentially determined by the gravitation and radiation pressure by the Sun. The trajectories of particles, subject to the gravitational force, are completely solved with the equations of motion, but it is not the case with the radiation pressure \citep{Bishop1989}.

The radiation pressure disturbs the conics (ellipses or hyperbolas) described by the particles under the influence of gravity. The resonant scattering of solar photons leads to a total momentum transfer from the photon to the atom or molecule \citep{Burns1979}. In the non-relativistic case, assuming an isotropic reemission of the solar photon, this one is absorbed in the Sun direction and scattered with the same probability in all directions. For a sufficient flux of photons in the absorption wavelength range, the reemission in average does not induce any momentum transfer from the atom/molecule to the photon. The momentum variation, each second, between before and after the scattering imparts a force, the radiation pressure. 

\citet{Bishop1989} analyzed its effect on the structure of planetary exospheres. Nevertheless, their work was limited only to the Sun-planet axis, with a null component assumed for the angular momentum around the Sun-planet axis. We thus generalize here their work to a full 3D calculation, in order to investigate the influence of the radiation pressure on the trajectories \citep{Beth2015a}, as well as the density profiles \citep{Beth2015b} and escape flux (future work).

In this paper, we propose to investigate the influence of the radiation pressure on the stability of planetary exospheres, based on dynamical effects of the radiation pressure. In particular, we study the validity of our approach (see. \citeauthor{Beth2015a}(\citeyear{Beth2015a}, \citeyear{Beth2015b}) for the specific case of Hot Jupiters, which needs to consider the Circular Restricted Three Body Problem (CR3BP) combined with the radiation pressure i.e. the photogravitational Three Body Problem. We will also derive a new expression of the Hill's sphere radius (the limit of the gravitationnal influence of the central body) that includes the effect of the radiation pressure, before we discuss the impact of the radiation pressure force on the stability of the Hot Jupiters atmospheres (with a case study on HD 209458b). These may be indeed be bounded or not to the planet depending on the radiation pressure intensity.

In the section \ref{formalism}, we develop our approach to study the photogravitational CR3BP. Then, in the section \ref{results}, we will present the topology of equipotentials, derive a new expression of the Hill's sphere and then discuss the possible consequence for the planetary atmosphere stability and the escaping flux. Finally, the conclusions will be reported in the section \ref{conclusions}.

\section{Formalism}\label{formalism}

We investigate here the combined effects of the external forces on planetary neutral exospheres: the planetary gravity, the stellar gravity, the centrifugal force and the radiation pressure. As we know, the radiation pressure is directly proportional to the flux of photons. Neglecting relativistic effects such as the Doppler shift or the Poynting-Robertson effect, the radiation pressure then depends only on the square of the distance from the star as the stellar gravity. We define the $\beta$ parameter as the ratio between the acceleration induced by the radiation pressure and by the stellar gravity, this parameter being a constant in the planetary System. Indeed, the radiation pressure effect without relativistic effects depends only on the flux and thus the square of the stellar distance as the stellar gravity.

Each neutral is then subject to the dimensionless effective potential $-\Omega(x,y,z)$ in the rotating coordinate system defined as:

\begin{equation}
\Omega(x,y,z)=\dfrac{1}{2}(x^2+y^2)+\dfrac{(1-\beta)(1-\mu)}{r_S}+\dfrac{\mu}{r_{pl}}
\end{equation}
and
\begin{equation*}
\left\{
\begin{array}{lcl}
r_S&=&\sqrt{(x+\mu)^2+y^2+z^2}\\
r_{pl}&=&\sqrt{(x+\mu-1)^2+y^2+z^2}\\
\end{array}
\right.
\end{equation*}
with $\mu=M_1/(M_1+M_2)$, $M_1$ the mass of the planet, $M_2$ the mass of the star, $(-\mu,0,0)$ the position of the star and $(1-\mu,0,0)$ the one of the planet. The axis are oriented such as the $x$-axis is the star-planet axis, the $z$-axis is perpendicular to the planetary motion and the $y$-axis so that the $Oxyz$ frame is a direct orthonormal basis. The notations are similar to the ones used in \citet{Simmons1985}.

The first term of $\Omega$ describes the potential energy associated to the centrifugal force, the second one corresponds to the stellar gravity and the radiation pressure effect $\beta$ and the last one the planetary gravity.

This problem was previously studied concerning the positions and the stability of equilibrium points by \citeauthor{Schuerman1972} (\citeyear{Schuerman1972},\citeyear{Schuerman1980}) and \citet{Simmons1985}. \citet{Schuerman1972} determined the position of the Lagrange points $L_4$, $L_5$, then investigated the equipotentials and the $\beta$ values for which the Roche lobe of the planet is connected to the stellar one through $L_1$ as a function of $\mu$ and $\beta$. For very small $\beta$ values, both are not connected at all (cf. \citet{Schuerman1972}, figure 4). On another hand, \citet{Simmons1985} investigated the general case of a binary star system, where each star has its own radiation pressure. They gave a complete review about the Lagrange points $L_1$, $L_2$ and $L_3$ with the numerical derivation of their positions and their stability, whereas $L_4$ and $L_5$ were mostly studied by \citet{Schuerman1980}.

The link between the stability of a planetary atmosphere and the CR3BP was tackled by \citet{Lecavelier2004} about the strong escape rate observed on HD 209458b and widely covered in the literature (\citet{Vidal2004},\citet{Lecavelier2004},\citet{Koskinen2013b}). They modelled with different densities and temperatures estimations  the upper atmosphere of HD 209458b. For high thermospheric temperatures, the thermosphere could extend so far that the exobase could reach the Roche lobe limit, located at only $\sim 4$ planetary radii. Thus, in this case the requisite kinetic energy to escape from the exobase (or critical radius $r_{c}$) is smaller than the common gravitational potential $GM_1/r_{c}$. Several papers dedicated to HD 209458b investigated this issue, focusing on the mass loss rate and the conversion of the EUV flux into kinetic energy for Hydrogen so that it can escape. However, most of the previous studies did not take into account the additional effect of the radiation pressure (i.e. $\beta=0$) which acts as a repulsive force from the star. 

%In this paper, we will thus study more specifically the topology of the equipotentials (or zero-velocity curves, ZVC) in the $(x,y)$-plane ($z=0$).

In this paper, we will discuss on the positions of the equilibrium points compared with the size of the extended planetary atmospheres and show how the radiation pressure will influence the stability of the atmospheres of Hot Jupiters, with using HD 209458b as a case study.

%Previous studies investigated the influence of the proximity of Lagrange points and its impact on atmospheric escapes. \citet{Lecavelier2004} were the first trying to study such an effect on the atmospheric structure of HD 209458b. For high thermospheric temperatures, the exobase, defined as the transition layer between collisional part ($\text{\it{Kn}}\ll1$) and the quasi-collisionless part ($\text{\it{Kn}}\gg1$), becomes close to the Roche limit, the closed equipotential passing through the Lagrange point $L_1$.

\section{Results}\label{results}

In this section \ref{results}, we present our results on the topology (\ref{topology}) and on the decreasing size of the Roche lobe due to the radiation pressure. We derive an analytical formula to approximate the Hill's sphere size depending on the radiation pressure intensity (\ref{Hill1}). We will then discuss the implications of the modified Hill's sphere on the planetary atmosphere stability (\ref{stability}). This modification will significantly affect the giant planets close to their host star such as the Hot Jupiters. This section will provide quantitative results for specific situations including the case of HD 209458b (and for hydrogen atoms), but both the approach and the results can be essentially applied to any planetary atmosphere.

\subsection{Topology of equipotentials}\label{topology}

In this first part, we study the topology of equipotentials (or ZVC curves) modified by the radiation pressure with the example of HD 209458b.

\begin{figure}
	\centering
\includegraphics[width=.4\linewidth]{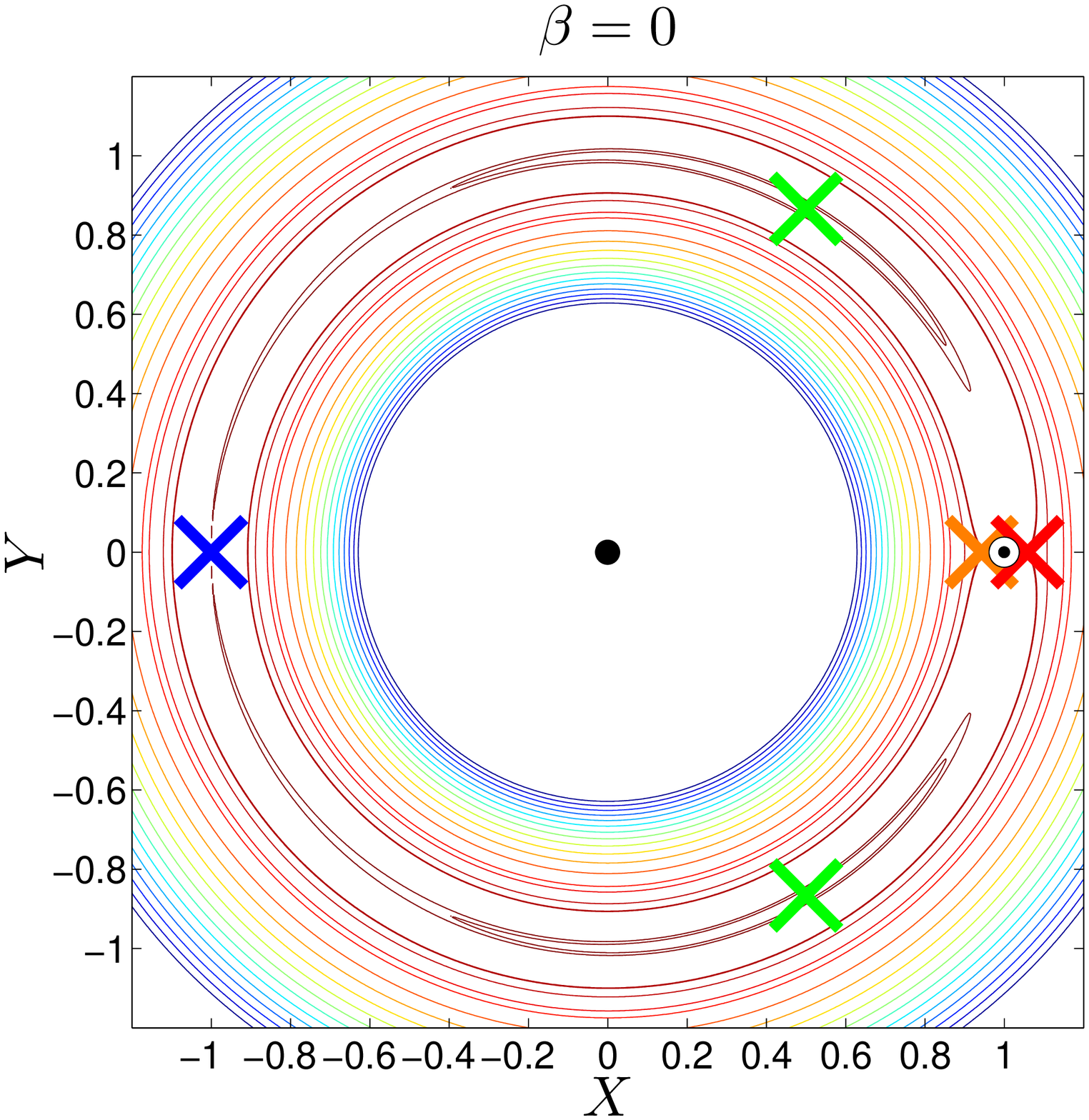}
\includegraphics[width=.4\linewidth]{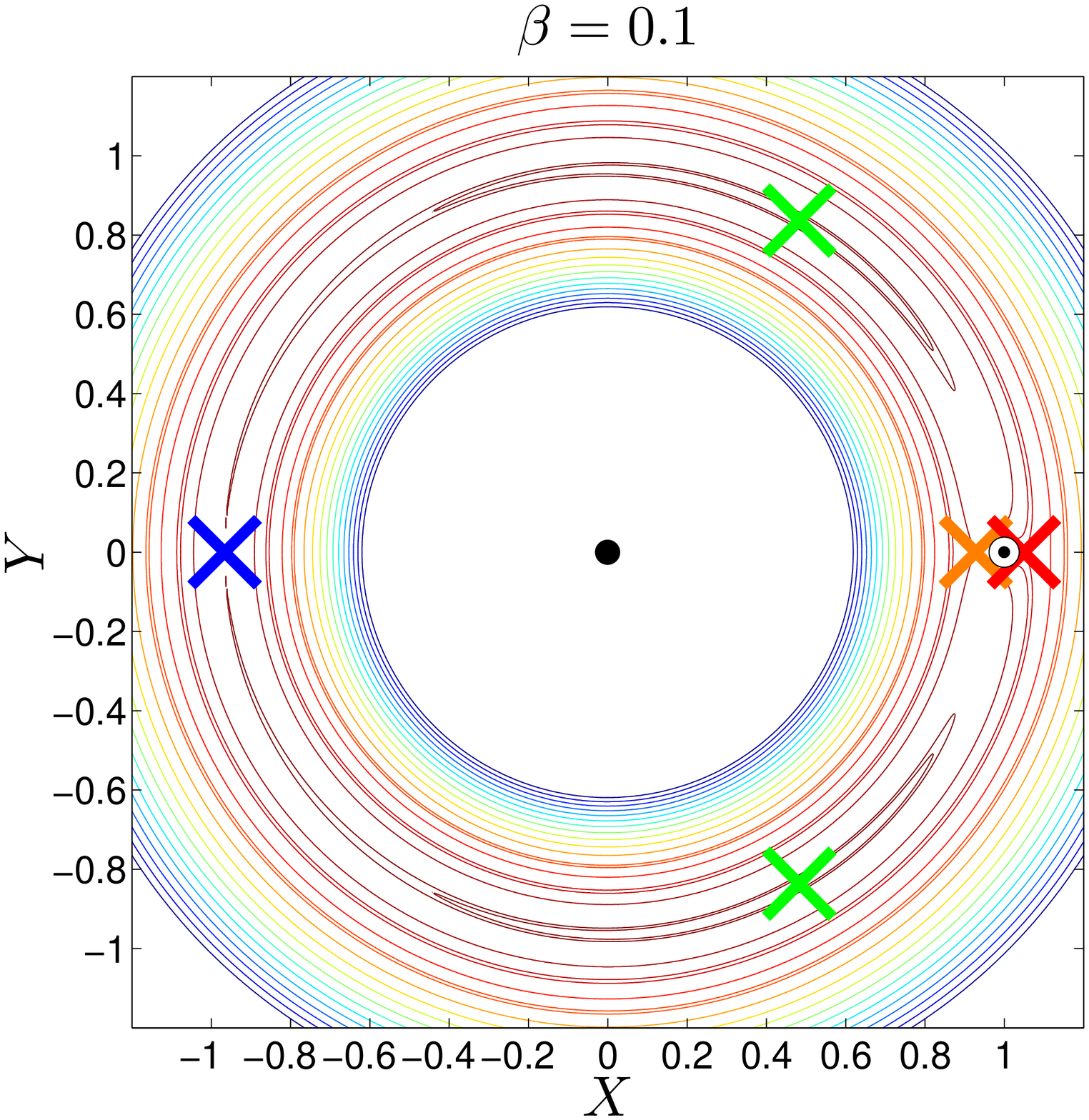}\\
\includegraphics[width=.4\linewidth]{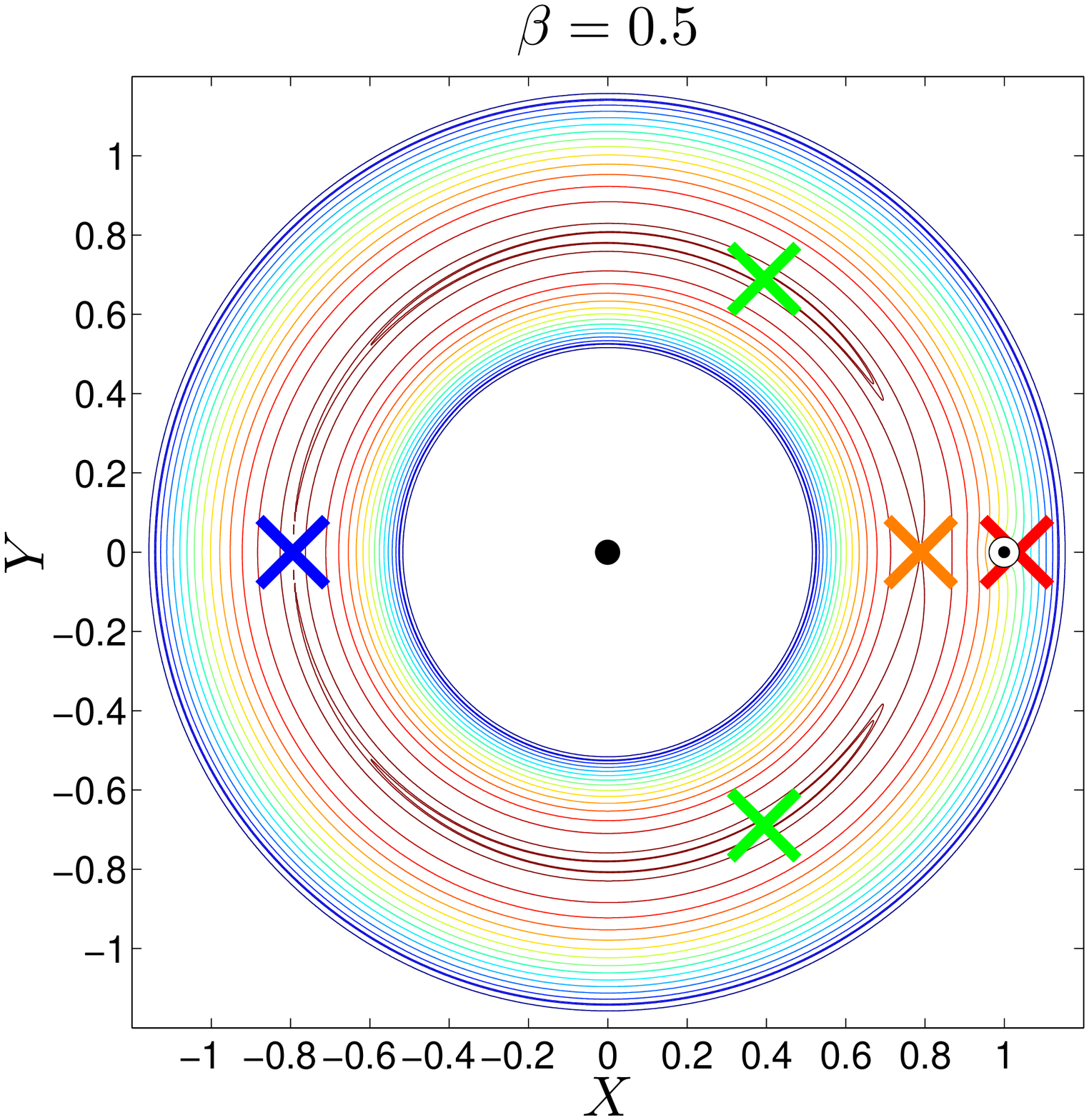}
\includegraphics[width=.4\linewidth]{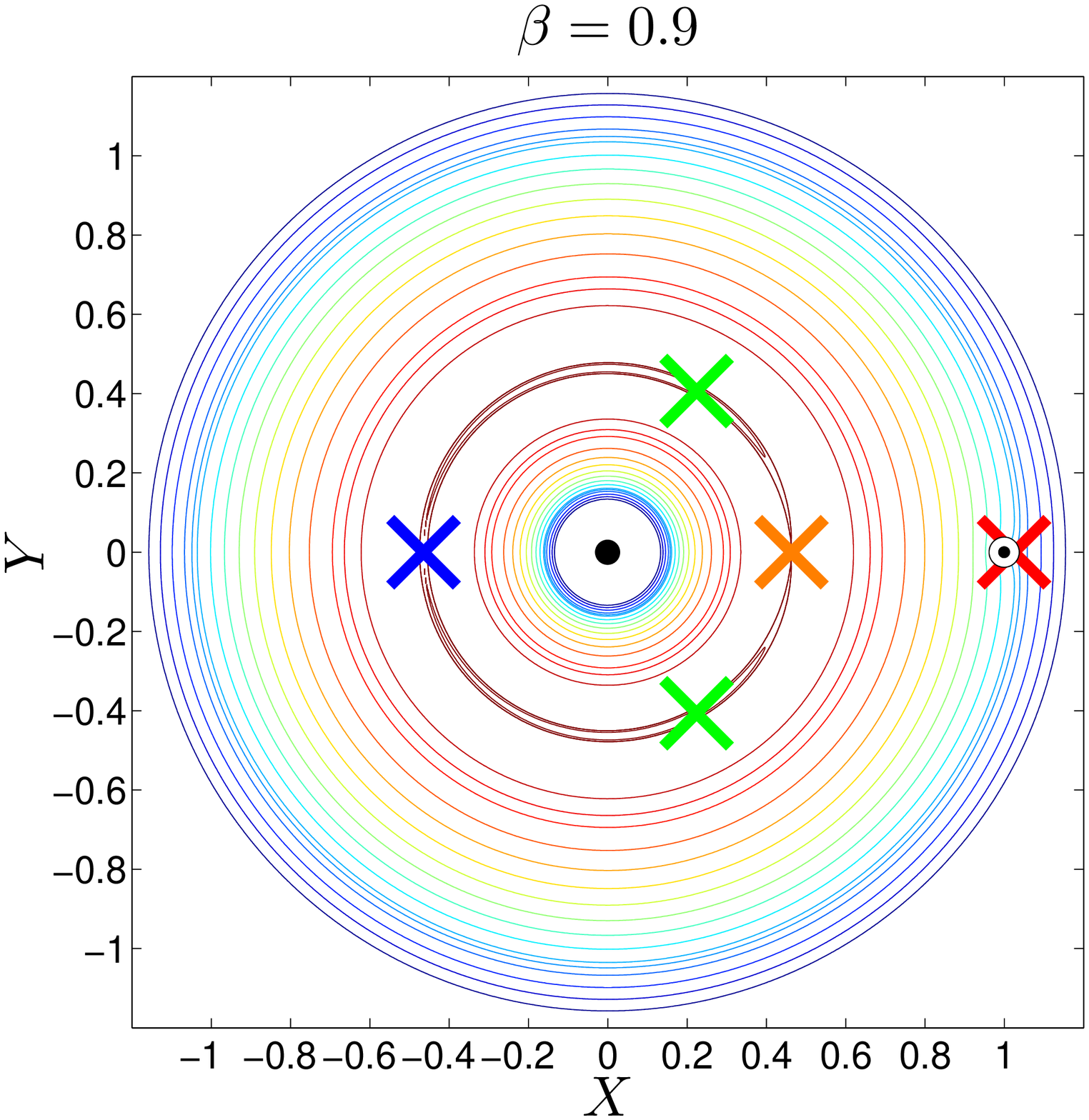}\\
\includegraphics[width=.4\linewidth]{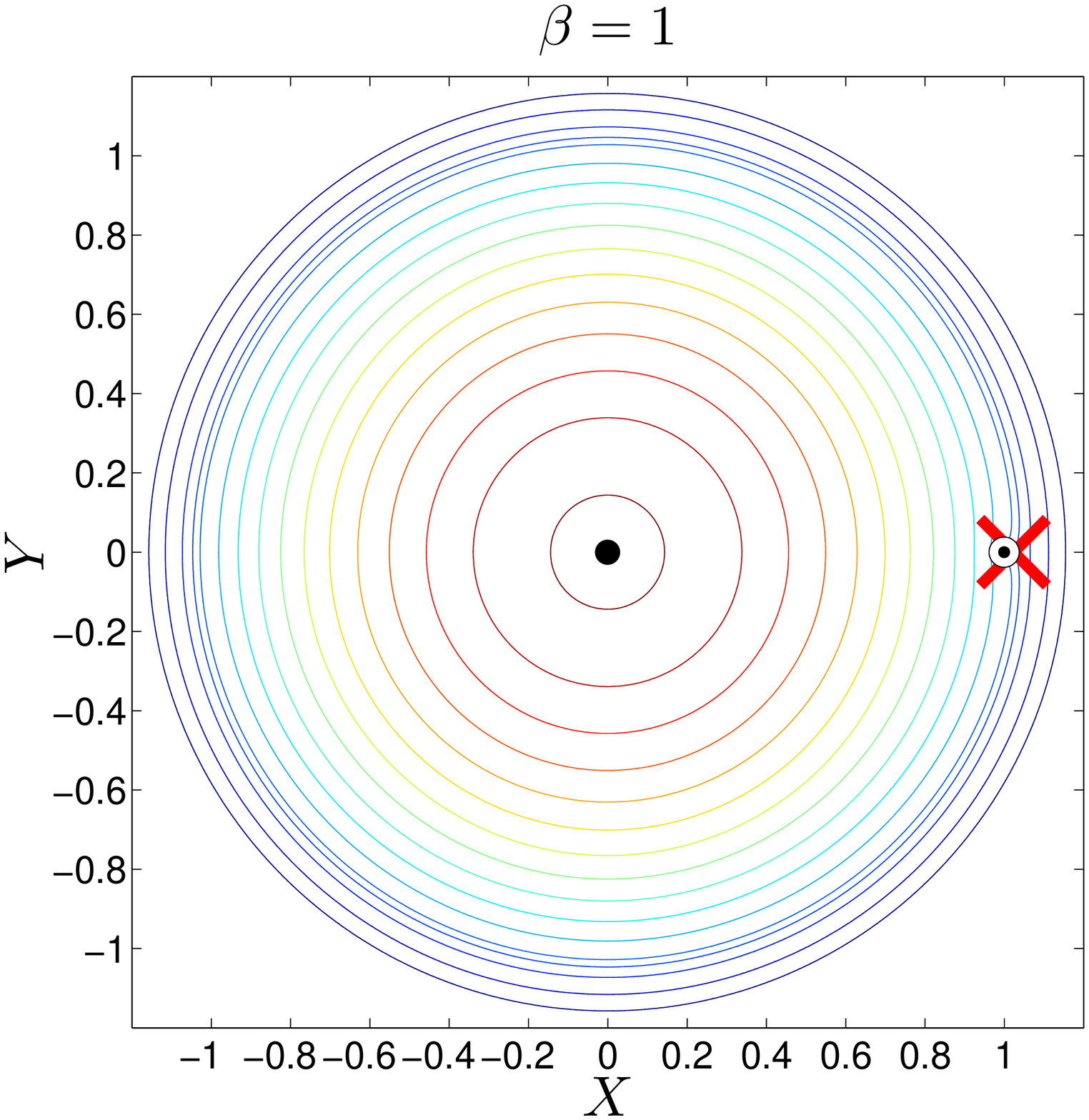}
\includegraphics[width=.4\linewidth]{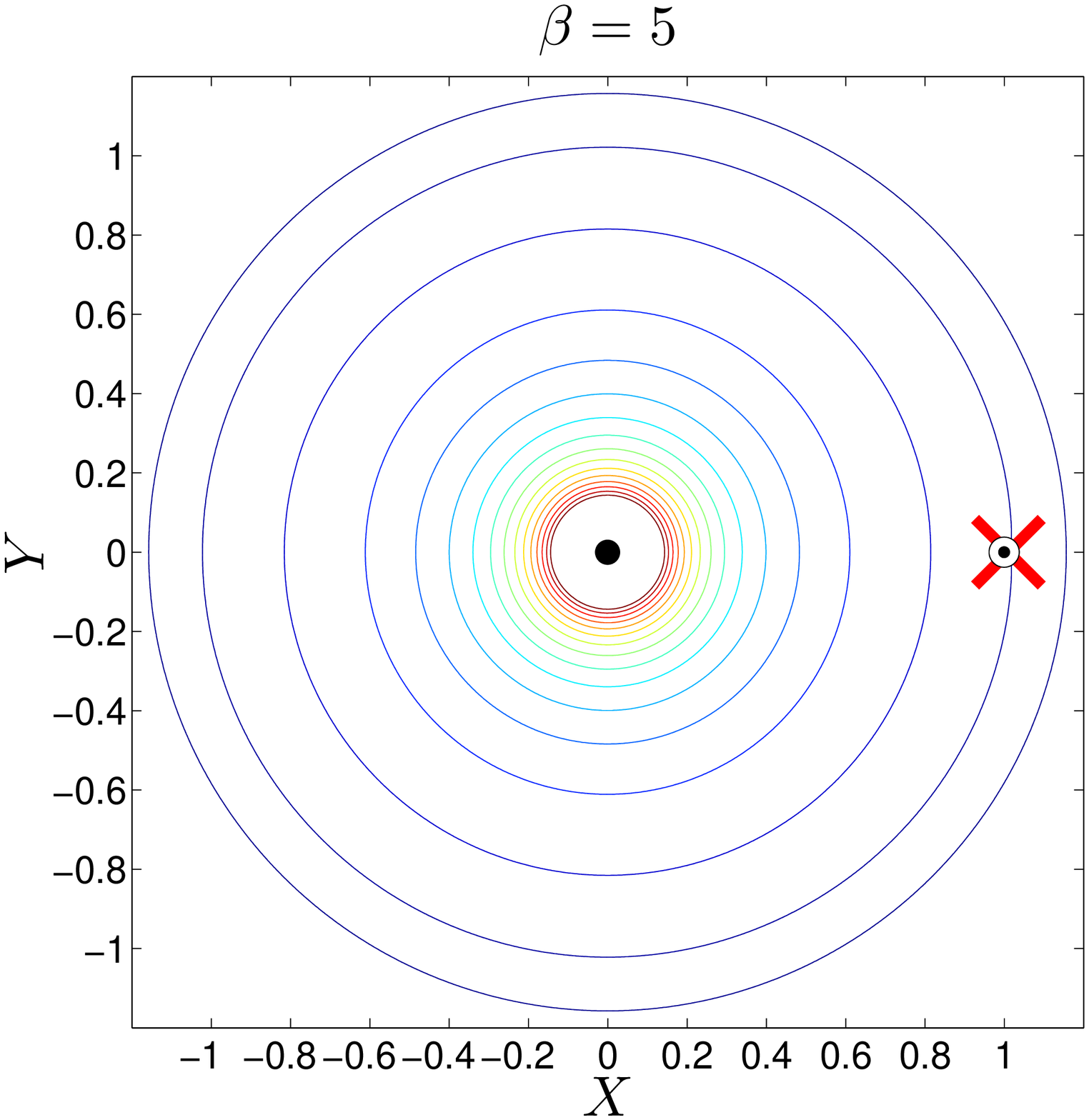}

\caption{Levels of effective potential $\Omega$ for $\beta=0$ (pure CR3BP, upper panels), $0.1$, $0.5$, $0.9$, $1$, $5$ for HD 209458b. The crosses define the Lagrange points: $L_1$ (orange), $L_2$ (red), $L_3$ (blue), $L_4$ and $L_5$ (green). The central point corresponds to the star position. The black point around $(1,0)$ corresponds to the true planetary size and the surrounding white disk to the atmosphere with its supposed size.}
\label{equipot}
\end{figure}

The figure \ref{equipot} shows the topology of the equipotentials for increasing $\beta$ values in the case of HD 209458b (i.e. $\mu\approx 5.83$, see \url{www.exoplanets.eu/HD209458b}). Extremely close to its star ($0.047$ A.U.), with a mass ($0.69$ $M_J$) and a size ($1.35$ $R_J$) of the same order as Jupiter's ones, HD 209458b is strongly affected by tidal forces \citep{Lecavelier2004}. Nevertheless, the photogravitational Three Body Problem was not yet covered concerning its implication on the stability of planetary exospheres; only the potential effects of tidal forces were investigated. We considered here a range of $\beta$ values of 0 to 5, to be compared with the ranges of $\beta$ values for Hydrogen at HD 209458b as reported by \citet{Bourrier2013} ( these authors take into account the Doppler shift so that $\beta$ depends on the velocity along the radial direction with respect to the star, $\beta(v\approx0)\approx4$).

As can seen in the figure \ref{equipot}, the different Lagrange points migrate in the direction toward the star. Indeed, the radiation pressure is a force opposed to the stellar gravity. For increasing $\beta$ values, the equilibrium points need to be nearer to the star in order to put up the radiation pressure acceleration.

For the critical value $\beta=1$, all Lagrangian points disappear except $L_2$ which should stay behind the planet. However, for any $\beta$, $L_2$ becomes closer to the planet, and thus the sphere of influence or the Roche lobe has a decreasing size as a function of $\beta$.

In order to estimate the influence of $\beta$ on the Roche lobe size, we derive in the next section a new Hill's sphere radius taking into account the radiation pressure.

\subsection{Hill's sphere}\label{Hill1}

We propose here to investigate the influence of the radiation pressure on the Roche lobe radius. As previously defined, the Roche lobe, in the case of $\beta=0$ (no radiation pressure), is the last equipotential closed around the planet and passing through $L_1$. On another side, the Hill's sphere radius is a first order approximation of the Lagrange $L_1$ and $L_2$ distances ($L_1$ and $L_2$ are approximatively at the same distance from the planet). Literally, the Roche lobe is inside the Hill's sphere. The radius of the Hill's sphere is approximately given at the first order by the value $\sqrt[3]{\mu/3}$ \citep{Hill1878} scaled by the star-planet distance.

Following the demonstration for the Hill's sphere radius $R_H$, if we take into account the radiation pressure (i.e. $\beta\neq0$), then $R_H$ - i.e the position of the $L_2$ Lagrange point - is given by the following relation:
\begin{equation*}
-(1-\beta)\dfrac{1-\mu}{(1+R_{H})^2}-\dfrac{\mu}{R_{H}^2}+(1-\mu+R_{H})=0
\label{Hill}
\end{equation*}

\begin{equation*}
-(1-\beta)(1-\mu)R_{H}^2-\mu(1+R_{H})^2+R_{H}^2(1+R_{H})^2(1-\mu+R_{H})= 0
\end{equation*}

\begin{equation}
R_{H}^5+(3-\mu)R_{H}^4+(3-2\mu)R_{H}^3+[\beta(1-\mu)-\mu]R_{H}^2-2\mu R_{H}-\mu=0
\label{exact}
\end{equation}

as presented in \citet{Simmons1985} (cf. equation 10 with $\delta_{1}^{3}=1-\beta$ and $\delta_{2}^{3}=1$ in their paper).

According to Descartes' rule \citep{Descartes1637}, this polynomial has only one positive root (because the sign between two successive monomials changes once), the Hill's sphere radius.
As we know, this polynomial does not have explicit roots (for polynomials with a higher degree than 4, the roots cannot be written explicitly as a function of coefficients, cf. Abel-Ruffini theorem \citep{Abel1824}). We can however reasonably assume that $R_{H}\ll 1$, $\mu \ll 1 $ and $\mu \ll R_{H}$: according to Hill's formula (without radiation pressure), $R_{H}(\beta=0)=O(\mu^{1/3})$; and according to \citep{Beth2015b}, a large radiation pressure leads to an exopause or Hill sphere radius located at $R_{H}(\beta)\approx(\mu/\beta)^{1/2}$. We thus cut this relation at the first power not depending on $\beta$ which is $R_{H}^3$ (for the initial derivation of the Hill's sphere radius, only the terms $R_{H}^3$ and $O(1)$ are kept). Thus, the relation \ref{Hill} gives us:
\begin{equation}
3R_{H}^3+\beta R_{H}^2-\mu\approx 0
\label{approximation}
\end{equation}

This polynomial has one positive root and two negative (if $\beta>9\sqrt[3]{\mu/12}$) or complex conjugates (if $\beta<9\sqrt[3]{\mu/12}$) roots.
The value of the generalized Hill's sphere radius is the positive root and can be found analytically thanks to the Cardano method:
\begin{equation}
R_{H}=\dfrac{\beta}{9}\left(2\ \cosh\left(\dfrac{1}{3}\ \argcosh\left(\dfrac{1}{2}\left(\dfrac{9R_{H0}}{\beta}\right)^3 -1\right)\right)-1\right)
\end{equation}
for $\beta<3\sqrt[3]{9\mu/4}$
\begin{equation}
R_{H}=\dfrac{\beta}{9}\left(2\ \cos\left(\dfrac{1}{3}\ \arccos\left(\dfrac{1}{2}\left(\dfrac{9R_{H0}}{\beta}\right)^3 -1\right)\right)-1\right)
\end{equation}
for $\beta>3\sqrt[3]{9\mu/4}$ with $R_{H0}=\sqrt[3]{\mu/3}$ the well known Hill's sphere radius without radiation pressure.\newline

\begin{figure}
	\centering
\includegraphics[width=.7\linewidth]{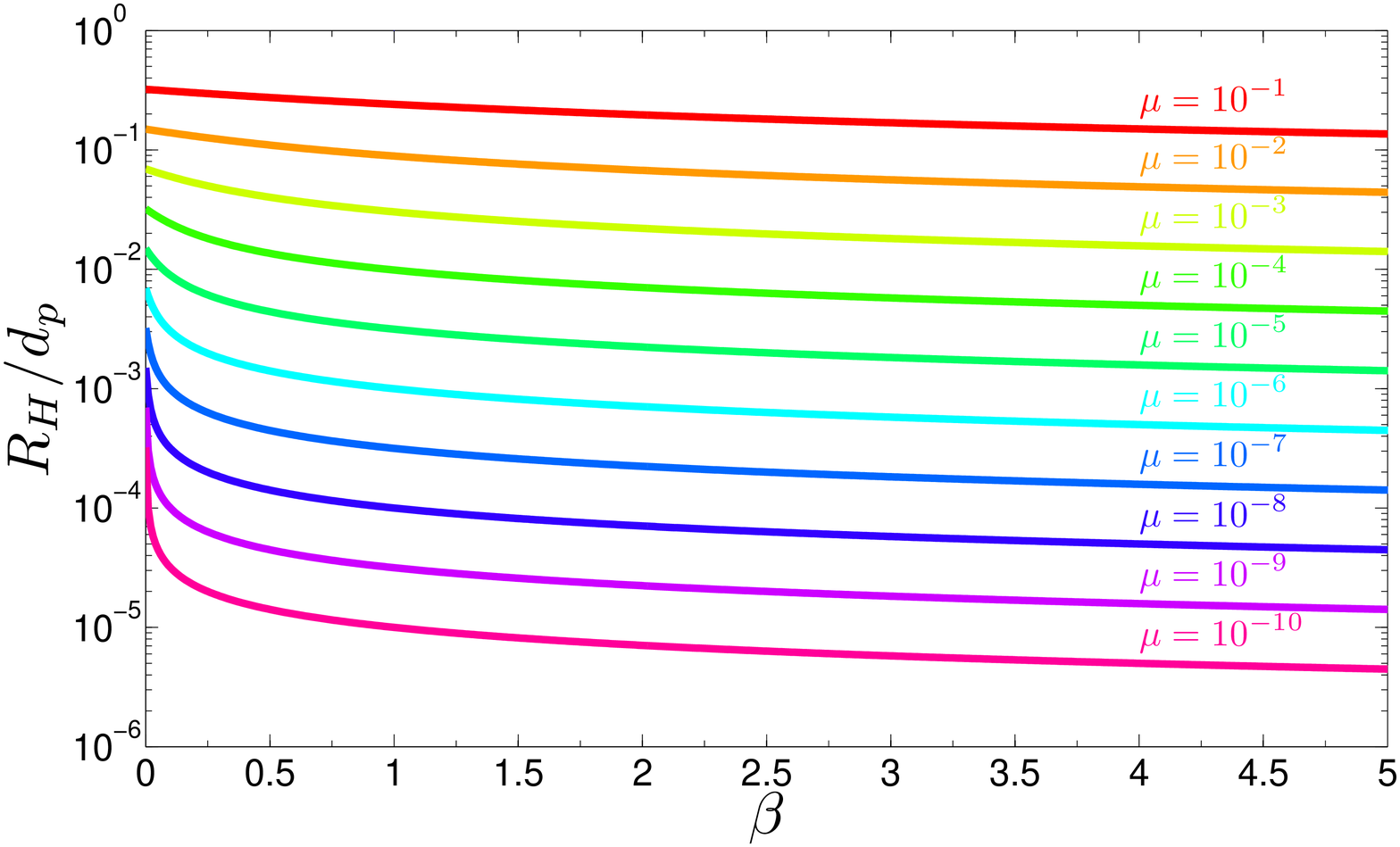}\\
\includegraphics[width=.7\linewidth]{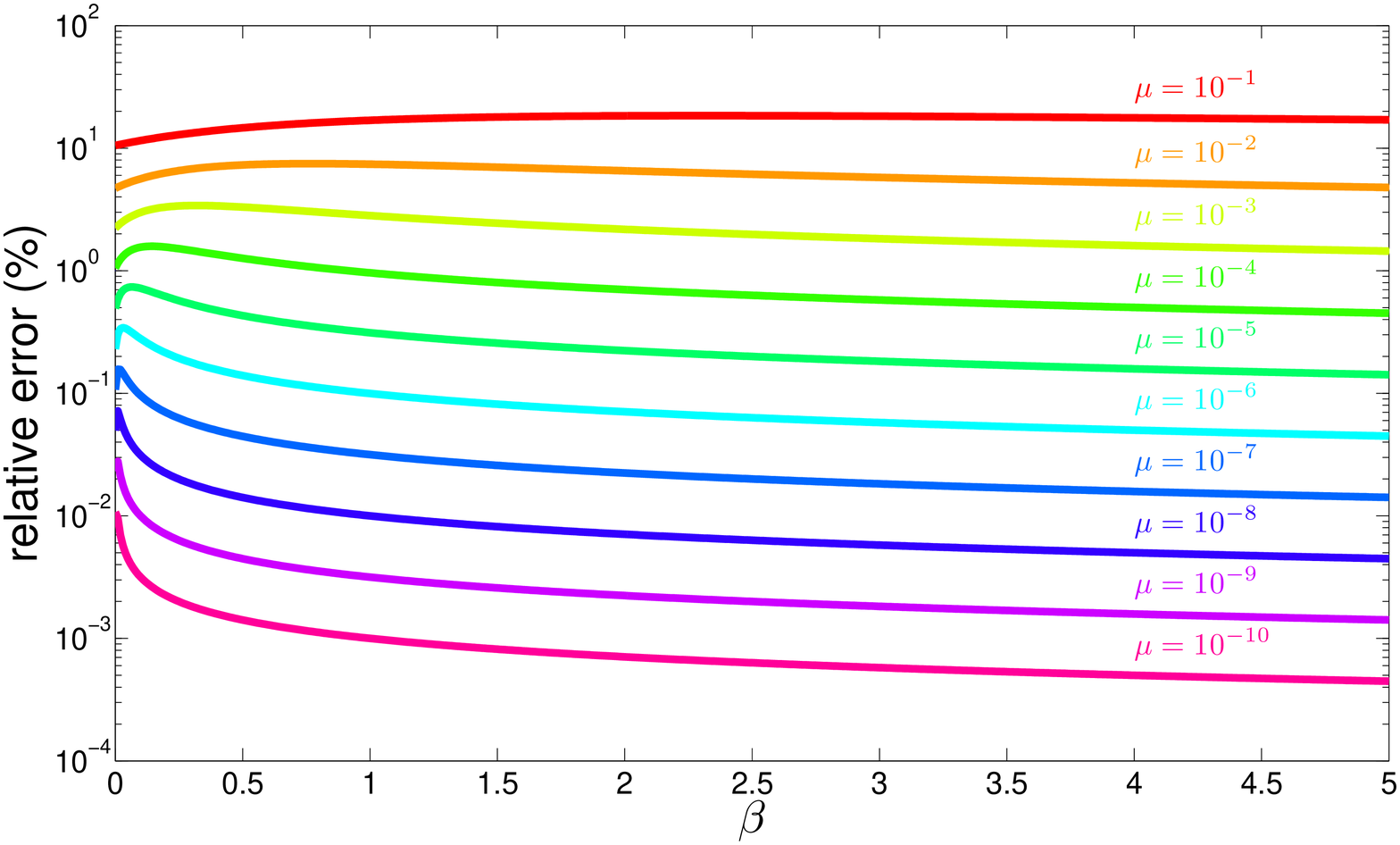}\\
\includegraphics[width=.7\linewidth]{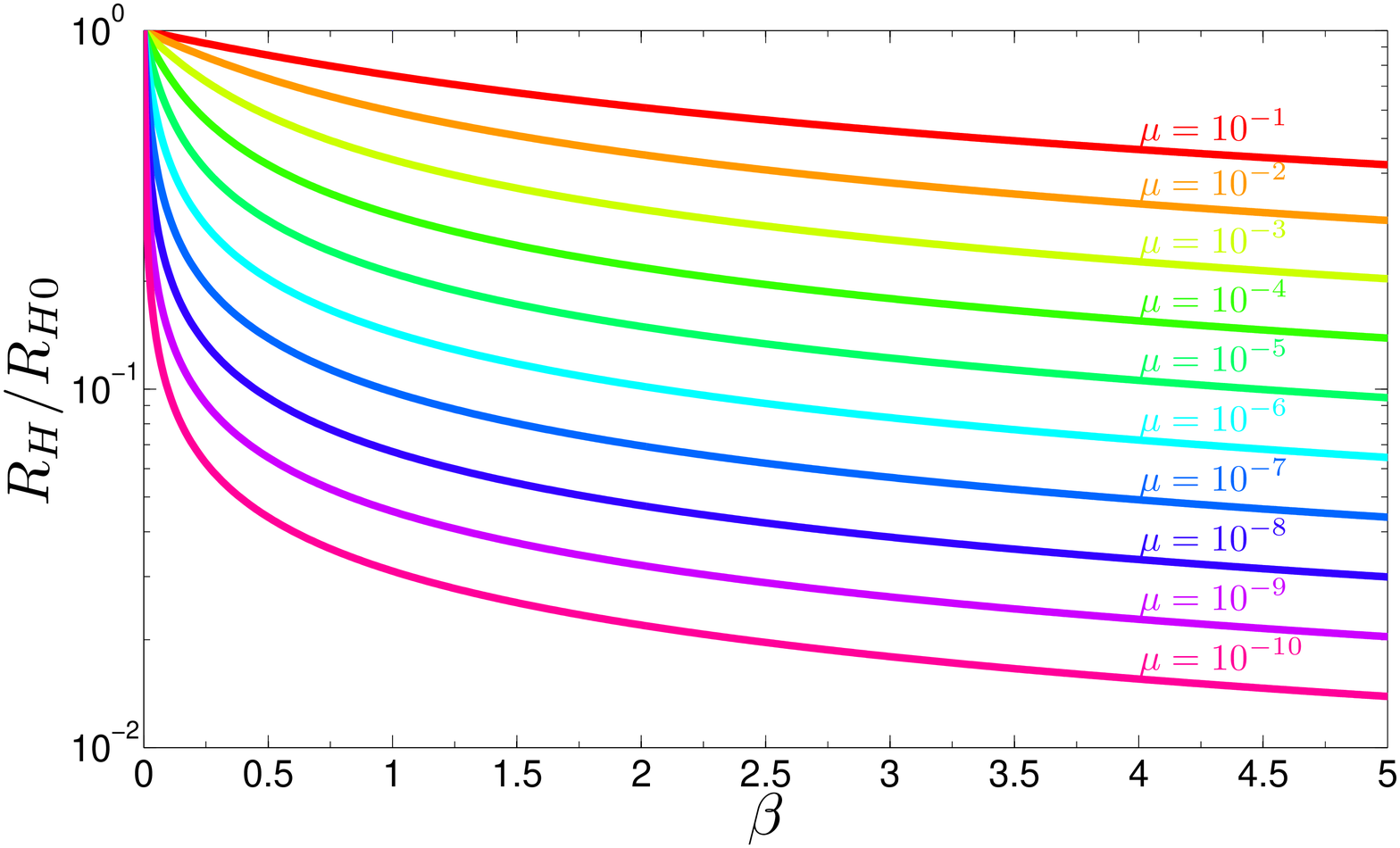}
\caption{(Upper panel) Ratio between the generalized Hill's sphere radius and the planet-star distance as a function of $\beta$. (Middle panel) Relative difference between the exact numerical root from the equation \ref{exact} and our analytical solution from the equation \ref{approximation}. (Lower panel) Same as upper panel, with the modified Hill's sphere radius  scaled with the Hill's sphere radius for $\beta=0$ (pure CR3BP), for different $\mu$ values. For $\beta=0$, we have the usual value of the Hill's sphere radius $\sqrt[3]{\mu/3}$ without radiation pressure. For higher $\beta$, the modified Hill's sphere radius converges asymptotically to $\sqrt{\mu/\beta}$, i.e. the exopause distance discussed by \citet{Beth2015b}.}
\label{RHill}
\end{figure}

The upper panel of figure \ref{RHill} shows the evolution of the Hill's sphere size as a function of $\mu$ and $\beta$. The planetary systems with low mass planet/star mass ratios are thus more sensitive to the radiation pressure (in particular at low beta values , i.e. $0$ to $0.5$), with a Hill's sphere radius decreasing more rapidly when the radiation pressure ($\beta$) increases.

To justify the use of this convenient formula, we compare our analytical solution of the equation \ref{approximation} with the solution obtained numerically from the equation \ref{exact} (see figure \ref{RHill}, middle panel). Our analytical formula thus approximates the position of $L_2$ with only $\sim3\%$ error for $\mu=10^{-3}$ (case of the Sun-Jupiter system). Moreover, this error decreases quickly with $\mu$ decreasing.

The Taylor series of our analytical formula gives at $\beta=0$ $\sqrt[3]{\mu/3}$ and at $\beta\longrightarrow+\infty$  $\sqrt{\mu/\beta}$. The $\beta=0$ asymptote corresponds to the classical Hill's sphere radius without radiation pressure, whereas the $\beta\longrightarrow+\infty$ asymptote corresponds to the exopause distance induced by the radiation pressure and discussed in detail by \citet{Beth2015b}. Our generalized Hill's sphere radius thus provides the location of the Hill's sphere radius for any regime of the stellar radiation pressure. Moreover, this gives a mathematical basis for the radiation pressure induced exopause first introduced by \citet{Bishop1991} and later numerically demonstrated by \citet{Beth2015b}. The generalized formula is all the more interesting since the asymptotic values can be used for restricted conditions of the radiation pressure. For an error below $\sim1\%$ the first asymptote (i.e. the classical Hill's sphere radius) is valid until $\beta=0.1R_{H0}$ only, whereas the second asymptote (i.e. the exopause distance) is valid only above $\beta=36R_{H0}$.

\begin{table}[!h]
\begin{center}
\begin{tabular}{|c|c|c|c|c|c|c|}
\hline
\rule[-1ex]{0ex}{4ex}Planets&$\beta=0$ &$\beta=0.6$    &$\beta=1.2$&$\beta=2$&$\beta=4$&$\beta=40$  \\
\hline
\hline
\rule[-1.2ex]{0ex}{4ex}Venus&167.0&36.22&25.59&19.80&13.99&4.42\\
\rule[-1.2ex]{0ex}{4ex}Earth&234.9&52.68&37.23&28.82&20.37&6.44\\
\rule[-1.2ex]{0ex}{4ex}Mars&319.7&49.63&34.99&27.06&19.12&6.04\\
\rule[-1.2ex]{0ex}{4ex}HD 209458b&4.24&2.14&1.57&1.23&0.88&0.28\\
\hline
\end{tabular}
\end{center}
\caption{Modified Hill's sphere radius in planetary radii with the effect of the radiation pressure for different planets (Venus, Mars, Earth and HD 209458b) and for different $\beta$ values (0,0.5,2,4,40).  The classical Hill's sphere radius (without radiation pressure) corresponds to $\beta=0$.}
\label{summary}
\end{table}

The $\beta$ parameter depends both on the stellar photon flux at each wavelength and on the species considered. This work investigates only the radiation pressure effect on Hydrogen atoms but for future works, other species such as Helium should be investigated too. In the table \ref{summary}, we provide the modified Roche lobe size in planetary radii for Venus, Earth, Mars and HD 209458b for various typical $\beta$ values. In the current Solar System, for Hydrogen, $\beta$ is between $0.6$ and $1.2$ with daily variations (\citet{Vidal1975}). The first column (i.e. beta=0) references the Hill's sphere radius in planetary radii as used in literature. With the effect of the radiation pressure, the size of the Hill's sphere may decrease strongly depending on the mass, the distance to the star and the radiation pressure ($\beta$). For example, the size of the Hill's sphere decreases by a factor of $6.5/6.3/9.1$ for Venus/Earth/Mars when $\beta=1.2$ (maximum radiation pressure) is considered instead of $\beta=0$ (no radiation pressure). The case of HD 209458b is also quite interesting: according to \citet{Bourrier2013}, the $\beta$ value for HD 209458b should be $\sim4$, neglecting relativistic effects, and thus the Hill's sphere radius should be below the surface according to our study.

We can also investigate the conditions encountered by the early Solar System. The Lyman-$\alpha$ intensity was 20 times higher for the one million year old Sun than at present \citep{Lammerbook}, leading to an estimated $\beta$ for Hydrogen of about 40. If we assume the planets were in the same place as today, the sphere of influence should be much closer to the planetary surface, at only a few planetary radii. 

The consequences on the atmospheric stability and escape flux, in particular for the case study HD 209458b, will be more detailed in the next sections \ref{stability} and \ref{casestudy}.

\subsection{Consequences on the stability of planetary atmospheres\label{stability}}

\begin{figure}
	\centering
\includegraphics[width=.7\linewidth]{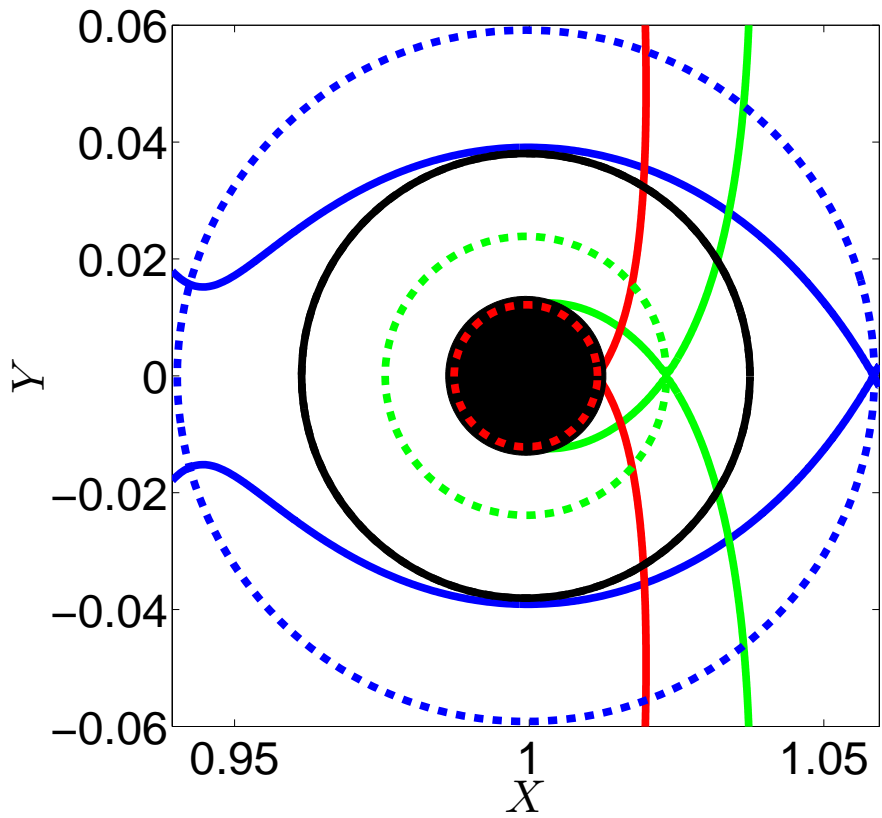}
\caption{Equipotentials around HD 209458b for different radiation pressure intensities: $\beta=0$ (blue), $1$ (green) and $4$ (red). The corresponding Hill's sphere is the circle in dotted line. The filled centered circle corresponds to the solid part of the planet and the surrounding circle to the exobase as proposed by \citet{Lecavelier2004}.\label{figure}}
\end{figure}

Regarding the example of the hot Jupiter HD 209458b, \citet{Lecavelier2004} proposed that the exobase of the planet is shaped by the Roche lobe proximity, since the $L_1$ Lagrange point is expected to be close to the planet due to the close host star. This then induces a geometrical blow-off: the exospheric particles need much less kinetic/thermal energy to escape by reaching the close Roche lobe. Based on our analysis, we agree in this case with the geometrical blow-off of HD 209458b proposed by \citet{Lecavelier2004} but for a different reason: as can be seen in figure \ref{figure}, the radiation pressure pushes the position of the Roche lobe (i.e. $L_1$) much closer to the star and thus much further from the planet, but the $L_2$ point is pushed closer to the planet below the expected exobase location (and even below the ``surface"), so that , no kinetic/thermal energy is needed for exospheric particles to escape even for a low exospheric temperature. 

As mentioned above the early Solar System planetary atmospheres encountered a Lyman-$\alpha$ flux about 20 times higher than today \citep{Lammerbook} and leading to a $\beta$ value of $40$ instead of $0.6-1.2$. Consequently, assuming the distance of Venus, Earth and Mars were the same as today, the sphere of influence for Hydrogen should be of the order of $4$ and $6$ planetary radii. The radiation pressure thus strongly limited the possible extension of the atmosphere: above this limit, the Hydrogen (and species with similar $\beta$) could not be bounded to the planet by the gravity and easily escaped to the interplanetary medium. This radiation pressure effect is a new constraint for understanding the planetary atmospheres evolution in the Solar or exoplanetary systems as well as for their modelling: the size of the simulation box should thus be at least larger than the Hill's sphere radius of each species to appropriately take into account the radiation pressure effect.

In the aim to have a complete picture of the atmospheric evolution of planets, which includes the influence of the radiation pressure, one should not only take into account the evolution of the stellar EUV-XUV fluxes that drive the heating/expansion of the atmospheres, but also the part of the flux spectrum that drives the radiation pressure. Depending on the flux spectrum of the host star, the species inside the atmosphere will be more or less affected by the radiation pressure based on their specific $\beta$ value. For example, if we refer to \citet{Chamberlain1987} for the Solar System conditions, the $\beta$ parameter of Helium can be between $0$ or very small values (e.g. $0.06$ for He II) and $260$ depending on the wavelength. If several species are not equally sensitive to the stellar radiation pressure, this could then affect the dynamic of each species above the exobase and yield to a differential drag of the species in the upper atmosphere, or to a protection of (radiation pressure) sensitive species such as Hydrogen by less sensitive ones through collisions. This last process could help understanding how Hydrogen rich atmospheres can remain stable for long periods despite a blow-off type escape regime expected. These questions will be investigated in a future work.

Moreover, it can be reasonably assumed that the radiation pressure will have a different effect on planetary atmospheres according to the Knudsen number {\it{Kn}}:
\begin{itemize}
	\item for low {\it{Kn}} (collisional regime), i.e. the scale height is greater than the mean free path, the radiation pressure will increase the mechanical energy of each Hydrogen atom on short distances similar to the mean free path. Then, through collisions, this increase will be redistributed quickly to other species and transformed into heat. The escaping flux will be a function of the incoming UV flux and its heating efficiency,
	\item for high {\it{Kn}} (non collisional regime), i.e. the mean free path is higher than the scale height, the radiation pressure will affect only the dynamics of each individual species on time scale shorter than $1/\nu$, $\nu$ the collisional frequency. The escaping flux will be function of the radiation pressure and the local thermal speed.
\end{itemize}

\subsection{Escaping flux: the case study HD 209458b\label{casestudy}}

As detailed above (and seen in table \ref{summary}), the hot Jupiter HD 209458b has its Hill's sphere radius (or exopause) below the exobase and even below the surface due to the strong radiation pressure of the host star. Since the exopause is below the exobase, all particles at the exobase thus escape (see \citet{Beth2015b}). This means the escape velocity is ``virtually" reduced to $0$ and thus the Jeans' parameter $\lambda_c=v_{esc}^2/v_{th}^2$ ($v_{esc}$ and $v_{th}$ being the escape and thermal speed) tends to $0$. The critical Jeans' parameter for the hydrodynamical blow-off regime proposed by \citet{Hunten1982} is $2$, but with the additional effect of the radiation pressure, $\lambda_c$ will reach even lower values. The thermal escape flux may then be given by the Jeans' escape formula applied to the whole exobase with $\lambda_c=0$, which leads for Hydrogen at HD 209458b to the following mass loss rate:
\begin{equation}
\dot{M}=4\pi r_{exo}^2 m \mathcal{F}_J(\lambda_c=0)
=n_{exo}r_{exo}^{2}\sqrt{8\pi m_{H} k_B T_{exo}}\approx 5.10^{10}\text{g}.\text{s}^{-1}
\label{escapeloss}
\end{equation}
for a temperature $T_{exo}$ of $8000$ K, an exobase at a distance of $2.8$ planetary radii and a density at the exobase of the order of $10^{13}$ m$^{-3}$ (\citet{Lecavelier2004}, \citet{Bourrier2013}). This loss rate is in agreement with the estimates by \citet{Vidal2004}, i.e. above $10^{9}$ g.s$^{-1}$, and of the same order as recent models which predict a range of loss rates between $10^9$ and $10^{11}$ g.s$^{-1}$ (\citet{Koskinen2013b}). However, the exobase temperature is poorly known and could be much lower than the one used by \citet{Lecavelier2004} to match their model with the spectroscopic observations. Then, we have performed as well the same calculation with the effective temperature of HD 209458b which is $T_{eff}\sim 1400$ K (\citet{Charbonneau2000},\citet{Evans2015}). Assuming the same order of magnitude for the density at the exobase than previously assumed, we obtain $\dot{M}=2.10^{10}\ \text{g}.\text{s}^{-1}$, to be compared with the Jeans' escape, which is here of about $\dot{M}\sim20\ \text{g}.\text{s}^{-1}$.

We thus obtain similar escape rates to previous models, but with a different process, i.e. with a geometrical ``blow-off" due to the radiation pressure influence and not due to the Roche lobe proximity. Furthermore, for Hot Jupiters, the thermal escape or Jeans' escape is in the literature neglected but our study shows that the radiation pressure acceleration modifies the gravitational potential in such a way that the exospheric particles are not bounded any more and/or the Jeans' parameter tends to $0$. Thus, the thermal escape is strongly enhanced by the stellar radiation pressure. Finally, our study shows that extreme temperatures, compared with $T_{eff}$, are not required to be in agreement with the expected mass loss rates for HD 209458b.

\section{Conclusions} \label{conclusions}

In this paper, we show the effect of the radiation pressure on the Circular Restricted Three Body Problem (CR3BP). This problem was previously tackled for Celestial Mechanics purposes but rarely in the context of Planetary Science. We first discuss the influence of the radiation pressure on the positions of the Lagrange points, before we derive for the first time an analytical formula to approximate the Hill's sphere radius with the additional effect of the radiation pressure. We highlight the strong interest of this derivation for the case studies of exoplanets and in particular Hot Jupiters. Most of the papers dealing with the Hill's sphere (or Roche lobe) and the stability of Hot Jupiters atmospheres do not take account the effect of the radiation pressure on the CR3BP. We show in extreme cases such as for Hot Jupiters like HD 209458b that the Hill's sphere radius could be drastically reduced and even be located below the surface. The components of the upper atmosphere, which are sensitive to the radiation pressure, are thus not ``bounded" to the planet any more and then escape to the interplanetary medium. The resulting increased escape calculated for HD 209458b is in agreement with the observations. Moreover, we show that this Hot Jupiter is in a ``blow-off" type escape regime due to the radiation pressure of the host star, in contradiction with previous works that considered a blow-off regime due to the proximity of the Roche lobe (the Roche lobe being actually pushed toward the star by the radiation pressure). Thus, the radiation pressure can modify the planetary escape flux and strongly affect the atmospheric evolution of hot Jupiters atmospheres from their early age, depending on the evolution of the host star as well.

In a future work, we will investigate how the intense radiation pressure of the young Sun may have influenced the atmospheric escape and evolution of the inner Solar System planets, whose Hill's sphere radius was significantly reduced.

%We define an arbitrary variable to prevent issues with turning points in the $x$ or $y$ directions and solve the system:
%\begin{equation}
%\left\{
%\begin{array}{rcl}
%\dfrac{\mathrm{d}x}{\mathrm{d}\tau}&=&x-\dfrac{(1-\mu)(x+\mu)}{({(x+\mu)^2+y^2})^{3/2}}-(1-\beta)\dfrac{\mu(x+\mu-1)}{({(x+\mu-1)^2+y^2})^{3/2}}\\\\
%\dfrac{\mathrm{d}y}{\mathrm{d}\tau}&=&-y+\dfrac{(1-\mu)y}{({(x+\mu)^2+y^2})^{3/2}}+(1-\beta)\dfrac{\mu y}{({(x+\mu-1)^2+y^2})^{3/2}}\\
%\end{array}
%\right.
%\end{equation}
%with different initial conditions. The calculation is performed with a Runge-Kutta method in C with double precision for efficiency reasons.

\section*{Acknowledgments}

This work was supported by the Centre National d'Études Spatiales (CNES).

\end{document}